\documentclass[12pt]{article}
\usepackage{fullpage,citesort,epsfig,psfrag,graphics,amsbsy,amssymb}
\usepackage{caption}
\newcommand{\beq}{\begin{equation}}
\newcommand{\eeq}{\end{equation}}
\newcommand{\bea}{\begin{eqnarray}}
\newcommand{\eea}{\end{eqnarray}}
\newcommand{\bfs}{\boldsymbol}

\newcommand{\be}{\begin{equation}}
\newcommand{\ee}{\end{equation}}
\newcommand{\bq}{\begin{eqnarray}}
\newcommand{\eq}{\end{eqnarray}}
\newcommand{\ket}[1]{|#1\rangle}
\newcommand{\bra}[1]{\langle#1|}

\def\math{\mathsurround=0pt }
\def\leftrightarrowfill{$\math \mathord\leftarrow \mkern-6mu 
 \cleaders\hbox{$\mkern-2mu \mathord- \mkern-2mu$}\hfill
 \mkern-6mu \mathord\rightarrow$}
\def\overleftrightarrow#1{\vbox{\ialign{##\crcr
     \leftrightarrowfill\crcr\noalign{\kern-1pt\nointerlineskip}
     $\hfil\displaystyle{#1}\hfil$\crcr}}}

\newcommand{\VEV}[1]{\langle#1\rangle}

\let\l=\lambda

 \def\bd{\begin{document}} \def\ed{\end{document}}
\def\ds{\documentstyle} \let\fr=\frac \let\bl=\bigl \let\br=\bigr
\let\Br=\Bigr \let\Bl=\Bigl
\let\bm=\bibitem
\let\na=\nabla
\let\pa=\partial \let\ov=\overline
\def\ft#1#2{{\textstyle{{\scriptstyle #1}\over {\scriptstyle #2}}}}
\def\fft#1#2{{#1 \over #2}}
\def\vp{\varphi}
\def\sst#1{{\scriptscriptstyle #1}}
\def\oneone{\rlap 1\mkern4mu{\rm l}}
\def\td{\tilde}
\def\wtd{\widetilde}
\def\dalemb#1#2{{\vbox{\hrule height .#2pt
        \hbox{\vrule width.#2pt height#1pt \kern#1pt
                \vrule width.#2pt}
        \hrule height.#2pt}}}
\def\square{\mathord{\dalemb{6.8}{7}\hbox{\hskip1pt}}}
\def\wtd{\widetilde}
\def\R{\rlap{\rm I}\mkern3mu{\rm R}}
\def\im{{\rm i}}
\def\tilg{\tilde{g}}
\def\tilF{\tilde{F}}
\def\tilA{\tilde{A}}
\def\varf{\varphi}
\def\tilf{\tilde{\phi}}
\def\tilh{\tilde{h}}
\def\rme{{\rm e}}
\def\ep{\epsilon}
\def\0{{(0)}}
\def\9{{(9)}}
\def\8{{(8)}}
\def\7{{(7)}}
\def\6{{(6)}}
\def\5{{(5)}}
\def\4{{(4)}}
\def\3{{(3)}}
\def\2{{(2)}}
\def\1{{(1)}}
\newcommand{\trace}{{\rm Tr}}
\newcommand{\ub}{\overline{U}}
\newcommand{\vb}{\overline{V}}
\newcommand{\uh}{\widehat{U}}
\newcommand{\vh}{\widehat{V}}
\newcommand{\ubh}{\overline{\widehat{U}}}
\newcommand{\vbh}{\overline{\widehat{V}}}
\newcommand{\lb}{\bar{\l}}
\newcommand{\Fb}{\overline{F}}
\newcommand{\Fh}{\widehat{F}}
\newcommand{\Fbh}{\overline{\widehat{F}}}
\newcommand{\Ab}{\overline{A}}
\newcommand{\Ah}{\widehat{A}}
\newcommand{\Abh}{\overline{\widehat{A}}}
\newcommand{\Gb}{\overline{G}}
\newcommand{\Gh}{\widehat{G}}
\newcommand{\Gbh}{\overline{\widehat{G}}}
\newcommand{\Pb}{\overline{P}}
\newcommand{\Ph}{\widehat{P}}
\newcommand{\Pbh}{\overline{\widehat{P}}}
\newcommand{\Qb}{\overline{Q}}
\newcommand{\Qh}{\widehat{Q}}
\newcommand{\Qbh}{\overline{\widehat{Q}}}
\newcommand{\Bb}{\overline{B}}
\newcommand{\Bh}{\widehat{B}}
\newcommand{\Bbh}{\overline{\widehat{B}}}
\newcommand{\fhns}{\hat{F}^{\rm (NS)}}
\newcommand{\fhrr}{\hat{F}^{\rm (RR)}}
\newcommand{\ahns}{\hat{A}^{\rm (NS)}}
\newcommand{\ahrr}{\hat{A}^{\rm (RR)}}
\newcommand{\hhrr}{\hat{H}^{\rm (RR)}}
\newcommand{\hchi}{\hat{\chi}}
\newcommand{\hphi}{\hat{\phi}}
\newcommand{\htau}{\hat{\tau}}
\newcommand{\cG}{{\cal G}}
\newcommand{\cGb}{\overline{{\cal G}}}
\newcommand{\cH}{{\cal H}}
\newcommand{\cP}{{\cal P}}
\newcommand{\cPb}{\overline{{\cal P}}}
\newcommand{\cQ}{{\cal Q}}
\newcommand{\cQb}{\overline{{\cal Q}}}
\newcommand{\cM}{{\cal M}}
\newcommand{\cN}{{\cal N}}
\newcommand{\cO}{{\cal O}}
\newcommand{\cD}{{\cal D}}
\newcommand{\cL}{{\cal L}}

\newcommand{\vpp}{\mbox{$\langle{\scriptstyle++}\rangle$}}
\newcommand{\vmp}{\mbox{$\langle{\scriptstyle-+}\rangle$}}
\newcommand{\vppp}{\mbox{$\langle{\scriptstyle+++}\rangle$}}
\newcommand{\vmpp}{\mbox{$\langle{\scriptstyle-++}\rangle$}}
\newcommand{\vpmp}{\mbox{$\langle{\scriptstyle+-+}\rangle$}}

\begin{document}
\setlength{\captionmargin}{36pt}
\begin{titlepage}
\begin{flushright}
UFIFT-HEP-09-05\\
\end{flushright}

\vskip 3cm
\begin{center}
\begin{large}
{\bf Summing Planar Open String Loops
on a Worldsheet Lattice\\ with Dirichlet
and Neumann Boundaries}
\end{large}

\vskip 2cm
{\large 
Charles B. Thorn\footnote{E-mail  address: {\tt thorn@phys.ufl.edu}}
}
\vskip0.8cm
{\it School of Natural Sciences, Institute for Advanced Study,
Princeton NJ 08540}
\vskip0.20cm
and
\vskip0.20cm
{\it Institute for Fundamental Theory\\
Department of Physics, University of Florida,
Gainesville FL 32611}


\vskip 1.0cm
\end{center}

\begin{abstract}
\noindent We extend the lightcone worldsheet lattice approach
to string theory, proposed in 1977 by Giles and me, 
to allow for coincident
D-branes. We find a convenient lattice representation 
of Dirichlet boundary conditions, which the open string 
coordinates transverse to the D-branes satisfy.  
We then represent
the sum over all planar open string multi-loop diagrams 
by introducing an Ising spin system on the worldsheet
lattice to keep track of the presence or absence of 
fluctuating boundaries.
Finally we discuss a simple mean field treatment of the resulting
coupled Ising/coordinate worldsheet system. The interplay between
Neumann and Dirichlet boundary conditions leads to
a richer phase structure, within this mean field approximation, 
than that found by Orland for the original system with 
only Neumann conditions.
\end{abstract}
\vfill
\end{titlepage}
\section{Introduction}
The problem of summing planar diagrams is central to
many issues in theoretical physics, from the
large $N$ approximation to QCD \cite{thooftlargen}
to the relationship of string theory to 
quantum field theory as exemplified by the
AdS/CFT correspondence \cite{maldacenasole}. In the
absence of an analytic solution to this problem,
it would be nice to have an approach amenable
to numerical simulations.
Over 30 years ago, Giles and I proposed a way
to digitize interacting bosonic string
theory, in its first quantized version, 
as a sum over histories on a lightcone worldsheet lattice
\cite{gilest} (GT). Although inspired by perturbation theory,
the resulting formalism provides a fully non-perturbative
dynamics which reproduces the formal\footnote{Technically 
Lorentz covariance in the
continuum limit requires counterterms to cancel lattice
artifacts that arise from boundary terms in the integration 
over moduli (see e.g.\cite{GNS,greenseiberg}). Bulk and
boundary terms already included in \cite{gilest} can 
account for some (perhaps all) of the necessary counterterms.} 
perturbation theory when expanded in 
powers of the string coupling $g$. The restriction of this
sum over histories to planar open string loops
should be manageable on a computer.

In recent years, my colleagues and I have constructed
an explicit lightcone worldsheet representation of the planar
diagrams of a wide range of matrix field theories 
\cite{bardakcit}. These
constructions rely on fermionic worldsheet
ghosts to cancel the bulk degrees of freedom of
each worldsheet coordinate -- the worldsheet systems
that reproduce field theory diagrams are essentially
topological. The ensuing negative
signs in the path integrands spell serious
difficulties for numerical
simulations of such worldsheet path integrals.
This problem can be avoided by replacing each 
field quantum with a finite tension open string, bringing
us back to the original GT formalism.
Thus we propose that numerical simulations of
planar diagram sums be performed
in the GT formalism at finite string tension,
after which conclusions about field theory can be drawn by
study of the infinite tension limit.

All open string
coordinates in \cite{gilest} satisfy 
free end (Neumann) boundary conditions.
It has long been clear that D-branes \cite{dailp} 
provide the key for arranging that
the infinite tension limit of critical string theory (in
26 or 10 space-time dimensions) yield
a quantum field theory in lower dimensional space-time.
D-branes are subspaces on which open strings end, meaning
that the open string coordinates describing motion perpendicular
to the D-branes satisfy Dirichlet boundary conditions.
The purpose of this article is to explain
how such conditions can be introduced in the 
GT worldsheet lattice formalism, and to begin to assess,
in the context of a simple mean field approximation,
their impact on the physics of planar diagram summation.

In addition to bringing in Dirichlet boundaries,
the GT lattice formalism also needs to
be extended to include Grassmann coordinates in order to
describe the Neveu-Schwarz (NS) boson model 
\cite{neveuschwarz,neveust}, 
Ramond fermions \cite{ramond,neveustfermions},
or the superstring \cite{gliozziso}. In particular,
we have suggested that the even G-parity Neveu-Schwarz 
open string model with SU($N$) Chan-Paton factors (NS$+$), 
which is free of open string
tachyons, could be used to establish
a string representation of 
large $N$ QCD \cite{thornsubqcd,thornnonabelian}.
Since the lightcone worldsheet lattice is tailor made
for summing planar diagrams, its extension to cover
the NS$+$ open string model would provide a promising
way to sum planar diagrams on a computer. Then
study of the $T_0\to\infty$ limit should yield new
information about large $N$ QCD. 

We conclude this introduction with a brief review of the
original GT lattice formalism for 
bosonic string theory \cite{gilest}.
It starts with a
lattice worldsheet path integral for the lightcone quantized 
free open string \cite{goddardgrt},
\bea
\bra{{\bfs x}_f}e^{-TP^-}\ket{{\bfs x}_i}
&\equiv&\int{\cal D}{\bfs x}
\exp\left\{-\int_0^Td\tau\int_0^{P^+}
d\sigma{1\over2}({\dot{\bfs x}}^2
+T_0^2{\bfs x}^{\prime2})\right\}\\
&\to& \int\prod_{j=1}^N\prod_{i=1}^Md{\bfs x}_i^j \exp\left\{
-{T_0\over2}\sum_{j=0}^N\sum_{i=1}^M({\bfs x}_i^{j+1}
-{\bfs x}_i^j)^2-V^{\rm open}_{\rm N}({\bfs x}_i^j)\right\}
\\
V_{\rm N}^{\rm open}&=&{T_0\over2}\sum_{j=1}^N\sum_{i=1}^{M-1}
({\bfs x}_{i+1}^j-{\bfs x}_i^j)^2
\eea 
where for simplicity we have taken the discrete unit of
$\sigma$ to be $T_0 a$ ($P^+=MaT_0$), with $a$ the discrete unit of $\tau$ ($T=(N+1)a$).
Then $a$ drops out of the formulae and the continuum limit
is simply $M,N\to\infty$ with $N/M=T_0T/P^+$ fixed.
In the above expression, ${\bfs x}_k^0\equiv {\bfs x}^i_k$
and ${\bfs x}_k^{N+1}\equiv {\bfs x}^f_k$ are fixed by
the initial and final states.

We have written the potential term $V_{\rm N}$ of the lattice action 
appropriate to the open string,
with Neumann boundary conditions which are automatic consequences
of the absence of a ``bond'' joining the sites $(1,j)$
to the respective sites $(M,j)$. The closed string
action would be obtained by simply restoring those bonds.
\bea
V^{\rm closed}({\bfs x}_i^j)&=&V^{\rm open}_{\rm N}({\bfs x}_i^j)
+{T_0\over2}\sum_{j=1}^N
({\bfs x}_{1}^j-{\bfs x}_M^j)^2
\eea
More generally, by rearranging the bond patterns in the
potential energy  term $V$ of the worldsheet action $S$, 
we can describe any number of
closed and open strings. Then interactions among 
strings can be achieved by summing over histories in which
the bond patterns change from time to time \cite{gilest}.
Each appearance or disappearance of a bond is accompanied
by a factor of $g$, and each bond interchange by a factor of
$g^2$.

This general sum over histories would involve wildly nonlocal 
interactions on the worldsheet, which would surely defeat any attempt
at numerical evaluation. However, the numerical prospects are much
brighter for the sum over histories corresponding to planar
open string multiloop diagrams\footnote{Incorporating an
SU($N$) ``color'' symmetry via Chan-Paton factors, these planar
diagrams are singled out by 't Hooft's large $N$ limit
\cite{thooftlargen}. With a canonically normalized gauge
coupling $g_s$ the limit holds $g_s^2N$ fixed. To simplify writing
we will absorb a factor of $\sqrt{N}$ in our coupling:
$g=g_s\sqrt{N}$}. 
In this case the only changes
in bond patterns would be the appearance or disappearance
of bonds between nearest neighbor sites. Since these
changes are all local on the worldsheet, techniques of
condensed matter physics and quantum field theory should apply. 
For instance, Orland 
\cite{orland} has applied the technique of mean field
theory to study the physics of this planar diagram summation\footnote{See \cite{bardakciopenmean} for a recent alternative treatment
of mean field theory in this context.}.
He introduced an Ising spin variable $s_i^j=\pm1$ to represent the
two states ``on'' ($s=+1$) or ``off'' ($s=-1$) of each planar bond.  
Putting $P_i^j=(1+s_i^j)/2$, we then have
\bea
\bra{{\bfs x}_f}e^{-TP^-}\ket{{\bfs x}_i}_{\rm Planar}
&\to& \prod_{i,j}
\sum_{P_i^j=0,1}\int\prod_{j=1}^N\prod_{i=1}^Md{\bfs x}_i^j 
\exp\left\{-S({\bfs x},P)\right\}\\
S({\bfs x},P)&=&
\sum_{j=0}^N\sum_{i=1}^M\Bigg[{T_0\over2}({\bfs x}_i^{j+1}
-{\bfs x}_i^j)^2+{T_0\over2}
P_i^j({\bfs x}_{i+1}^j-{\bfs x}_i^j)^2\nonumber\\&&\hskip1in 
+\alpha+\beta(1-P_i^j)-(P_i^{j+1}
-P_i^j)^2\ln g\Bigg]
\label{planarsum}
\eea
The terms on the last line account for the coupling constant
$g$ and the bulk ($\alpha$) and boundary ($\beta$)
worldsheet counterterms.
Since we have included a fluctuating bond between $i=1$
and $i=M$, this expression describes the planar evolution
of a closed string. For the corresponding evolution of
an open string, one simply imposes the constraint $P_M^j=0$.

Although (\ref{planarsum}) is completely
well-defined and finite for fixed $M,N$, it 
allows for only the two natural counterterms that
were introduced in \cite{gilest}, and shown
to be necessary
at the level of free strings ($g=0$). Indeed,
the lattice evaluation predicts that the closed and open free 
bosonic string 
ground state energies have the $M\to\infty$ behavior
\bea
P^-_{closed}&\sim& {D-2\over a}\left[{2MG\over\pi}\right]
-{(D-2)\pi T_0\over6P^+}\\
P^-_{open}&\sim& {D-2\over a}\left[{2MG\over\pi}
-{1\over2}\ln(1+\sqrt{2})\right]-{(D-2)\pi T_0\over24P^+}
\eea
where $G=\sum_{n=0}^\infty(-)^n/(2n+1)^2\approx 0.9159656$ 
is Catalan's constant.
In \cite{gilest} we observed that the divergent (and non-Lorentz invariant)  
terms can be absorbed in $\alpha$ and $\beta$. Choosing
\bea
\alpha = -{2G(D-2)\over\pi} + O(g^2),
\qquad \beta={D-2\over2}\ln(1+\sqrt{2})
+O(g^2)\eea
gives finite, Lorentz covariant, and correct values for the
free string energy spectrum. Although it was not clearly
stated in \cite{gilest}, we must expect that both of
these counterterm parameters will receive corrections for nonzero
coupling $g\neq0$ in order to maintain Lorentz covariance
for $D=26$. Indeed, the one loop corrections in bosonic
string theory do contribute to them. An important open
problem, not addressed in this article,
is to resolve whether these two counterterms suffice
to render the loop expansion covariant. If not, any
further counterterms must be identified and incorporated into
the formalism.

The rest of the article is organized as follows. In Section
2 we present our prescription for handling Dirichlet
boundary conditions on the lattice. In Section 3, we 
represent the sum over planar diagrams, in which some open string
coordinates satisfy Neumann boundary conditions and others
satisfy Dirichlet boundary conditions, as a sum over
Ising spin configurations, where the Ising spin keeps
track of the fluctuating boundaries.  Then in Section 4,
we give our implementation of the mean field approximation
to the Ising spin dynamics. Section 5 concludes the
article with a discussion of our results and problems for the
future.
\section{Dirichlet Conditions on the Worldsheet Lattice} 
For notational clarity we shall use ${\bfs x}$ to
describe the string coordinates satisfying Neumann conditions, and 
we shall use ${\bfs y}$ for the string coordinates satisfying 
Dirichlet conditions. 
We first consider a single free string with Dirichlet conditions
${\bfs y}_0^j={\bfs y}_M^j=0$. The simplest way to discretize this 
string is to use the action:
\bea
S^\prime&=&{T_0\over2}\sum_{j=0}^N\sum_{i=1}^{M-1}({\bfs y}_i^{j+1}
-{\bfs y}_i^j)^2+{T_0\over2}\sum_{j=1}^N\left(\sum_{i=1}^{M-2}
({\bfs y}_{i+1}^j-{\bfs y}_i^j)^2+({\bfs y}_1^{j})^2
+({\bfs y}_{M-1}^j)^2
\right)
\eea
Note that $S^\prime$ involves only $M-1$ integration variables for
each time slice $j$.
Now consider how to obtain this action from the closed string action,
which involves $M$ integrations for each $j$, as would be necessary
in the sum over histories.
Then, the replacement
\bea
({\bfs y}_1^j-{\bfs y}_M^j)^2 + ({\bfs y}_M^j-{\bfs y}_{M-1}^j)^2
&\to& ({\bfs y}_1^{j})^2+({\bfs y}_{M-1}^j)^2,
\eea
encounters the problem that the coordinate ${\bfs y}_M$ 
describes a spurious zero frequency mode. An easy fix for this
is to include an extra term $T_0\sum_{j=1}^N({\bfs y}_M^j)^2$ in
$S_D$:
\bea
S_{\rm D}&\equiv&{T_0\over2}\sum_{j=0}^N\sum_{i=1}^M({\bfs y}_i^{j+1}
-{\bfs y}_i^j)^2+V_{\rm D}({\bfs y}_i^j)\nonumber\\
V_{\rm D}({\bfs y}_i^j)&=&{T_0\over2}\sum_{j=1}^N\left\{\sum_{i=1}^{M-2}
({\bfs y}_{i+1}^j-{\bfs y}_i^j)^2+{\bfs y}_1^{j2}
+{\bfs y}_{M-1}^{j2}+2{\bfs y}_M^{j2}
\right\}
\label{daction}
\eea
Then, since the frequency of this added mode is $O(1)$ in lattice
units, the mode described
by ${\bfs y}_M^j$ is irrelevant in the continuum limit. 
But retaining it allows
an efficient description of the creation and destruction of
Dirichlet boundaries without changes in the number of degrees of
freedom.

Before turning to that, we give the explicit evaluation of
the path history sum for the propagation of a free Dirichlet string. 
\newcommand{\sh}{{\rm sh}}
\newcommand{\ch}{{\rm ch}}
Define
\bea
\alpha_n&\equiv&4\sin^2{n\pi\over2(N+1)},\qquad n=1,2,\ldots,N\\
\beta_m&\equiv&4\sin^2{m\pi\over2M},\qquad m=0,1,\ldots,M-1
\eea
which are the respective eigenvalues of the kinetic and
potential bilinear forms occurring in $S_{\rm N}$. 
The eigenvalues of 
the potential bilinear form appearing in $S_{\rm D}$ are the
$\beta_m$, $m=1,\ldots,M-1$, plus the eigenvalue $2$
for the extra coordinate ${\bfs y}_M^j$. For economy of
writing it is convenient to put $\beta_M\equiv2$ and to define the
frequencies
\bea
\omega_m\equiv 2\sinh^{-1}{\sqrt{\beta_m}\over2}, 
\qquad{\rm for}~m=1,\ldots, M
\eea
Then the path integral for a Dirichlet string propagating
in $d$ dimensions over time $T=(N+1)a$
from ${\bfs y}_i=0$ to ${\bfs y}_f=0$ using (\ref{daction})
is
\bea
\bra{{\bfs 0}}e^{-(N+1)aP^-}\ket{{\bfs 0}}_D&=&
\left\{\left[{ T_0\over2\pi}\right]^{MN}\prod_{n=1}^N
\prod_{m=1}^{M}(\alpha_n+\beta_m)\right\}^{-d/2}\nonumber\\
&=&\left\{\left[{ T_0\over2\pi}\right]^{MN}\prod_{m=1}^{M}
{\sinh((N+1)\omega_m)
\over \sinh(\omega_m)}\right\}^{-d/2}\\
&\sim&e^{-(N+1)E_M}\left[{ T_0(1+\sqrt{2})\over2\pi}\right]^{Md/2}
{(3M)^{d/4}\over\prod_{m=1}^{\infty}(1-w^{2m})^{d/2}}\nonumber
\eea
where we have taken $M$ large in the last line and introduced 
$w\equiv e^{-(N+1){\pi/ M}}=e^{-T\pi T_0/P^+}$. Here,
\bea
E_M&=&{Md\over2}\ln{ T_0\over2\pi}+{d\over2}\sum_{m=1}^{M}\omega_m\\
&\sim&{Md\over2}\left({4G\over\pi}
+\ln{ T_0\over2\pi}\right)+{d\over2}\ln{2+\sqrt{3}\over1+\sqrt{2}}
-{\pi\over24M}+O(M^{-2})
\eea
The first two terms, one linear in $M$ and the other
independent of $M$, contribute divergent terms to the
continuum $P^-=E_M/a$ and violate Lorentz 
invariance. But they can be cancelled by the bulk and boundary
counterterms respectively. 
Here we see explicitly that the effect of the ${\bfs y}_M$
mode we added is simply to modify the coefficient $\beta$
of the boundary counterterm. Taking the case of $D-2$ 
Neumann and $26-D$ Dirichlet open string coordinates,
we see that we should have 
\bea
\beta&=&-{26-D\over2}\ln{2+\sqrt{3}\over1+\sqrt{2}}+{D-2\over2}
\ln(1+\sqrt{2})+O(g^2)\nonumber\\
&=&-{26-D\over2}\ln(2+\sqrt{3})+12\ln(1+\sqrt{2})+O(g^2)
\eea
For $D=4$ this is $\beta=-3.91+O(g^2)$. Curiously, at
$g=0$, $\beta$
stays negative for $D<10$ and is positive for $D\geq10$.
For $D=10$, $\beta\approx0.041+O(g^2)$.
\section{Summing Planar Diagrams}
We turn now to the problem of representing the sum of planar
diagrams by introducing the same system of Ising spins $s_i^j=\pm1$
or equivalently $P_i^j=(1+s_i^j)/2=0,1$ used in the case of 
Neumann conditions \cite{orland}. Associate each coordinate ${\bfs y}_i^j$
with the corresponding $P_i^j$ and let $P_i^j=0$ when
the Dirichlet condition applies to ${\bfs y}_i^j$. Then
we should write the potential term as:
\bea
&&\hskip-.5in{T_0\over2}\sum_{i,j}\left[P_i^jP_{i+1}^j
({\bfs y}_{i+1}^j-{\bfs y}_{i}^j)^2+(1-P_i^j)P_{i+1}^j
{\bfs y}_{i+1}^{j2}+(1-P_i^j)P_{i-1}^j
{\bfs y}_{i-1}^{j2}+2(1-P_i^j){\bfs y}_i^{j2}\right]\nonumber\\
&&=T_0\sum_{i,j}\left[{\bfs y}_i^{j2}-P_i^jP_{i+1}^j
{\bfs y}_i^j\cdot{\bfs y}_{i+1}^j\right].
\eea 
Then the lightcone worldsheet action that sums the planar
diagrams of Dirichlet open strings would be
\bea
S_{\rm D}&=&{T_0\over2}\sum_{j=0}^N\sum_{i=1}^M({\bfs y}_i^{j+1}
-{\bfs y}_i^j)^2+{T_0}\sum_{j=1}^N\sum_{i=1}^{M}
({\bfs y}_i^{j2}-{\bfs y}_i^j\cdot{\bfs y}_{i+1}^jP_i^jP_{i+1}^j)
\label{dplanarsum}
\eea
Notice that this implementation of Dirichlet conditions has
the feature that if {\it every} site is 
Dirichlet, i.e. $P_i^j=0$ for all $i,j$, then the system is
just $M$ independent oscillators with frequency of $O(1)$,
and the continuum limit would show no interesting physics.

In order to describe $D=4$ dimensional physics with a 
critical string theory in $26$ (bosonic) or $10$ (Neveu-Schwarz)
space-time dimensions, one can, as in the
development of the AdS/CFT correspondence  \cite{maldacenasole},
introduce a stack of $N$ coincident D3-branes, 
which are $3+1$ dimensional subspaces on which
open strings end. Let us call the 4 coordinates parallel
to the D3-branes $x^\mu$ and the coordinates perpendicular
to the D3-branes $y^I$. For the bosonic string $I$ takes
22 values and for the Neveu-Schwarz string it takes on 6
values. The coordinates $x(\sigma,\tau)$ for an open string
satisfy Neumann boundary conditions $\partial x/\partial\sigma=0$
whereas the coordinates $y$ satisfy Dirichlet boundary conditions
$y^I=0$. Then a possible worldsheet lattice set up would
be
\bea
S&=&\sum_{i,j}[\alpha+\beta(1-P_i^j)]
-\sum_{i,j}(P_i^{j+1}
-P_i^j)^2\ln g+{T_0\over2}\sum_{j=0}^N\sum_{i=1}^M[({\bfs x}_i^{j+1}
-{\bfs x}_i^j)^2+({\bfs y}_i^{j+1}
-{\bfs y}_i^j)^2]\nonumber\\
&&+{T_0\over2}\sum_{j=1}^N\sum_{i=1}^{M}
[({\bfs x}_{i+1}^j-{\bfs x}_i^j)^2P_i^j+2{\bfs y}_i^{j2}-2{\bfs y}_i^j\cdot{\bfs y}_{i+1}^jP_i^jP_{i+1}^j] 
\eea
The sum over planar diagrams is accomplished by summing over all
spin configurations $P_i^j=0,1$. It is worth pointing out the
physical situations represented by the extreme spin configurations.
If all $P_i^j=1$ The $x$'s and $y$'s appear on an equal footing and
represent a single closed string moving in 25 spatial dimensions (for
the bosonic string). In the opposite extreme, with all $P_i^j=0$,
the $x$'s see no potential and represent $M$ free 
Newtonian particles moving in 2 spatial dimensions, and the $y$'s
represent $M$ such particles bound by a harmonic oscillator
potential to the point ${\bfs y}=0$.

The parameters $\alpha$ and $\beta$ 
characterize the bulk and boundary 
counterterms respectively. They will depend on $g$ in a way that
we do not know {\it a priori}. It may well be that counterterms
beyond these will be required to ensure Lorentz invariance,
though there remains a slender hope that these will suffice.
A study of multi-loop corrections in perturbation theory will
be needed to resolve this issue. As a working hypothesis we shall
assume in this article that only these counterterms play a role.
\section{Mean Field Theory}
Mean field theory provides a simple method to understand
the physics of large systems, although it can be
misleading especially near critical points. A convenient
framework for applying mean field theory to our spin system 
begins with the addition
of a source term $\sum_{ij}\kappa_j^iP_i^j$ to the action
$S$. Then, writing the path integral in the presence of $\kappa$
as ${\cal Z}(\kappa)\equiv e^{-{\cal F}(\kappa)}$,
the expectations and correlators of the $P$'s can
be obtained as derivatives of ${\cal F}$ with 
respect to the $\kappa$'s. 
Defining
\bea
\phi_i^j\equiv \VEV{P_i^j} = {\partial {\cal F}\over\partial\kappa_j^i},
\eea
it follows that the Legendre transform 
${\cal A}(\phi)\equiv {\cal F}-\sum_{ij}\kappa_j^i\phi_i^j$ satisfies
\bea
{\partial {\cal A}\over\partial\phi_i^j}&=& -\kappa_j^i . 
\eea
Thus the possible values of $\phi$ are stationary points of 
the effective action ${\cal A}$ in the absence of sources.

\begin{figure}
\begin{center}
\psfrag{phi}{$\phi$}
\psfrag{V_s}{$\hskip-12pt{\cal V}_s$}
\includegraphics[width=2.75in,height=3in]{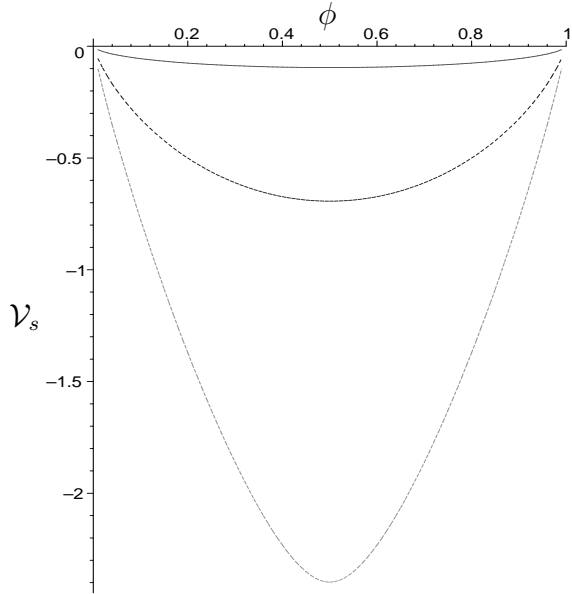}
\caption{The spin contribution to the effective potential, showing
the case of $\beta=0$ and $g=0.1$ (highest curve) $g=1$
(middle curve) and $g=10$ (lowest curve).}
\label{vsg0}
\end{center}
\end{figure}

\begin{figure}
\begin{center}
\psfrag{phi}{$\phi$}
\psfrag{V_x}{$\hskip-.3in {{\cal V}_x\over D-2}$}
\psfrag{V_y}{$\hskip-.35in {{\cal V}_y\over26-D}$}
\includegraphics[width=2.75in,height=3in]{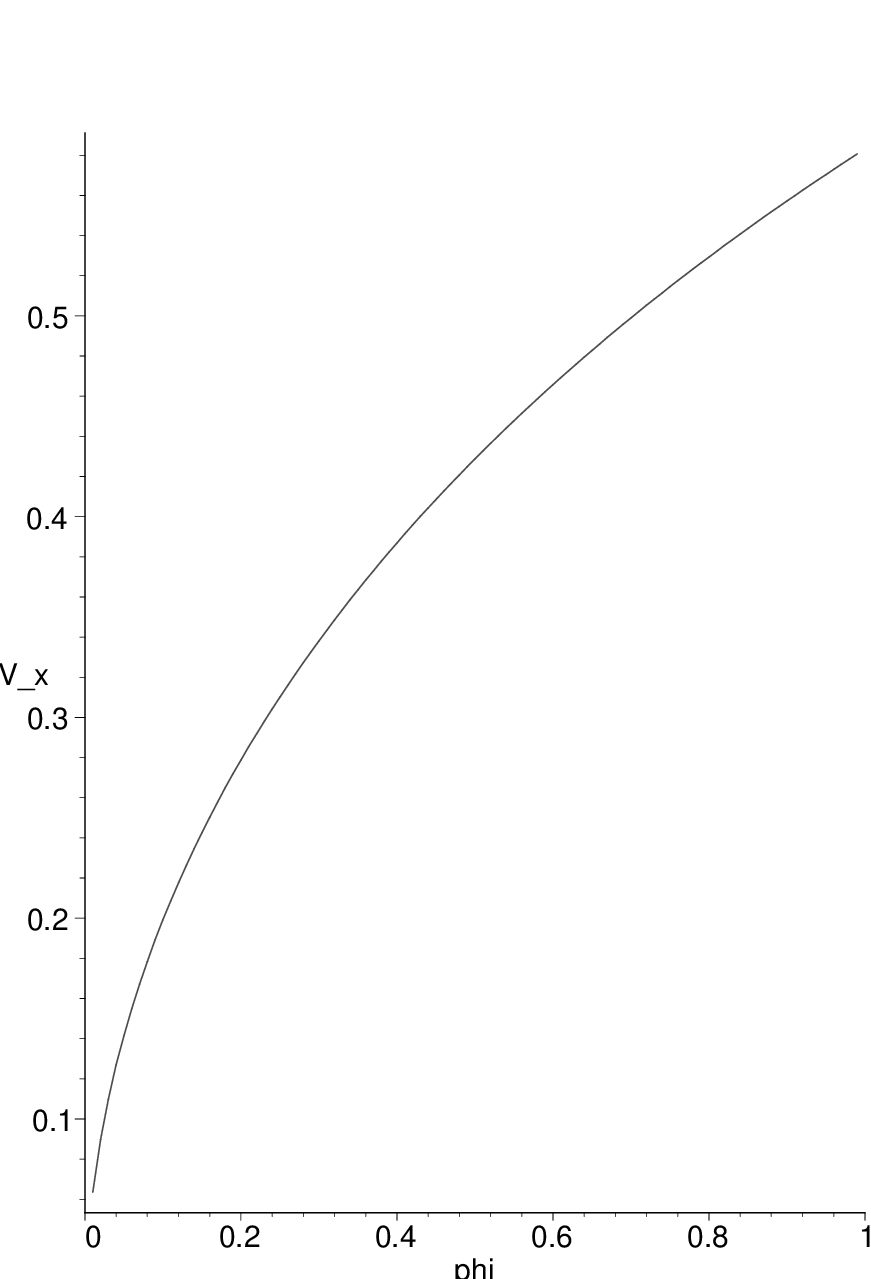}\qquad
\includegraphics[width=2.75in,height=3in]{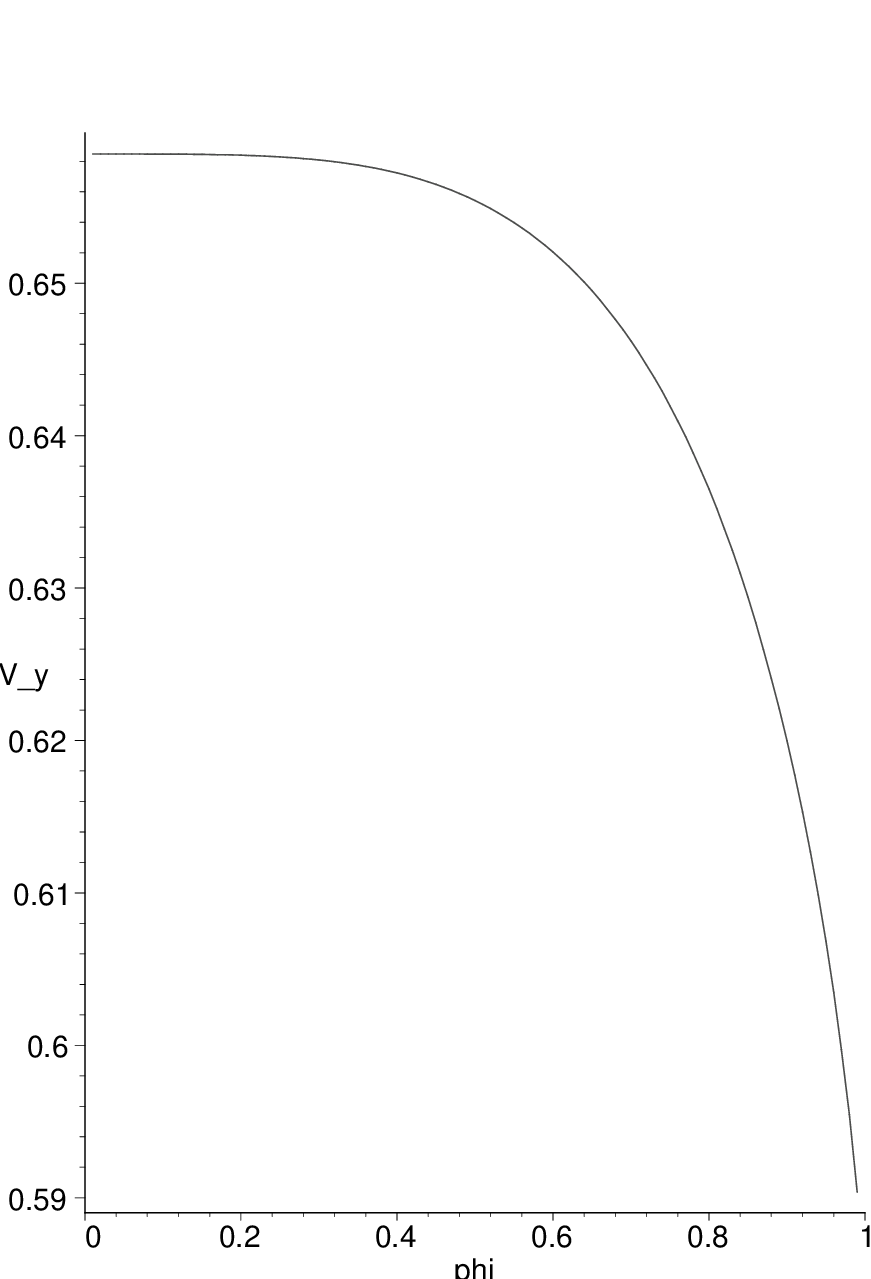}
\caption{The coordinate contributions to the effective potential,
Neumann on the left and Dirichlet on the right.}
\label{vxvy}
\end{center}
\end{figure}

\begin{figure}
\begin{center}
\psfrag{phi}{$\phi$}
\psfrag{V_xy}{$\hskip-.5in {\cal V}_x+{\cal V}_y$}
\includegraphics[width=4.5in,height=3.5in]{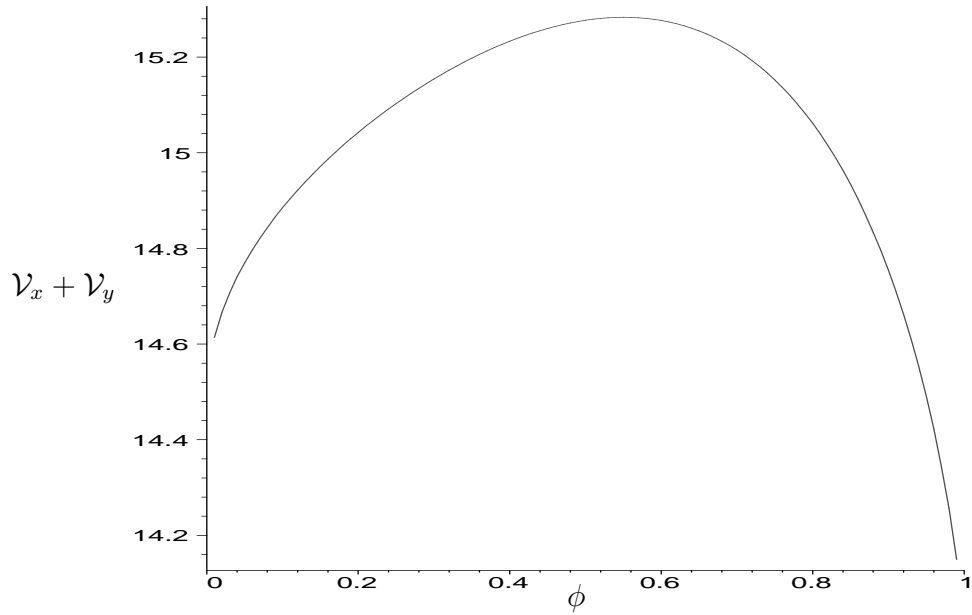}
\caption{The total coordinate contribution 
to the effective potential,
combined according to the case $D=4$.}
\label{vxy}
\end{center}
\end{figure}

\begin{figure}
\begin{center}
\psfrag{phi}{$\phi$}
\psfrag{A}{$\hskip-18pt{\cal V}$}
\includegraphics[width=2.75in,height=3in]{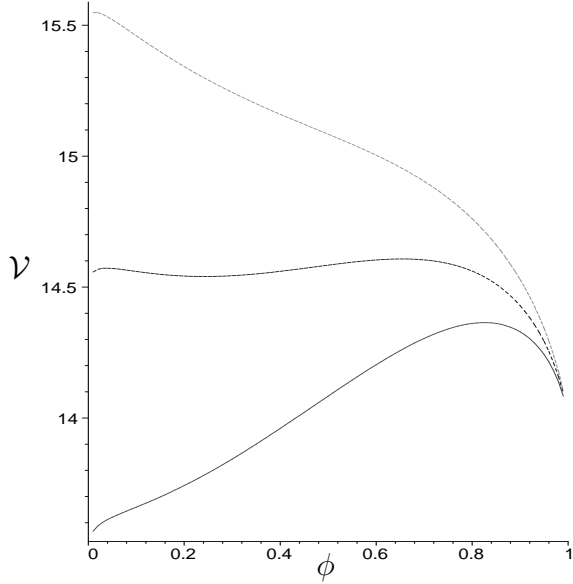}
\caption{The  effective potential, showing
the case of $g=1$ and $\beta=-1$ (lowest curve) $\beta=0$
(middle curve) and $\beta=1$ (highest curve).}
\label{a1beta}
\end{center}
\end{figure}

\begin{figure}
\begin{center}
\psfrag{phi}{$\phi$}
\psfrag{A}{$\hskip-18pt{\cal V}$}
\includegraphics[width=2.75in,height=3in]{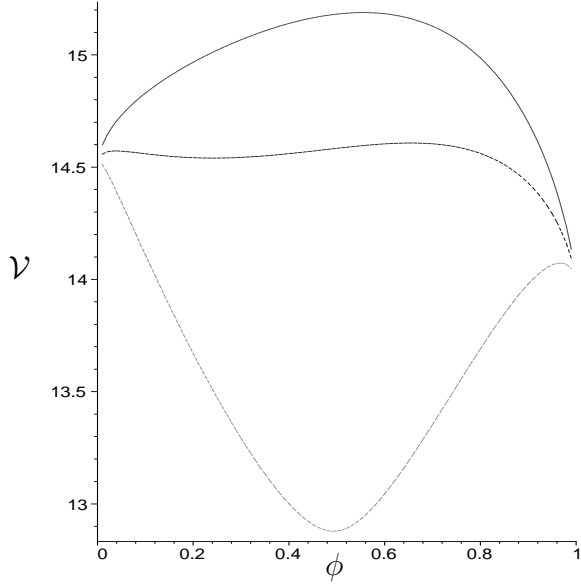}
\caption{The effective potential, showing
the case of $\beta=0$ and $g=0.1$ (highest curve) $g=1$
(middle curve) and $g=10$ (lowest curve).}
\label{ag0}
\end{center}
\end{figure}

Up to this point no approximations have been made. The mean
field approximation consists in replacing the coefficients
of the coordinate terms in the action by their expectation
values:
\bea
&&\hskip-12pt P_i^j\to 
\VEV{P_i^j}\ =\ \phi_i^j, \qquad P_i^jP_{i+1}^j
\ \to\ \VEV{P_i^jP_{i+1}^j}\ \equiv \phi_{2i}^j \\
&&\hskip-12pt S\to\sum_{i,j}[\alpha+\beta(1-P_i^j)]
-\sum_{i,j}(P_i^{j+1}
-P_i^j)^2\ln g+{T_0\over2}\sum_{j=0}^N\sum_{i=1}^M[({\bfs x}_i^{j+1}
-{\bfs x}_i^j)^2+({\bfs y}_i^{j+1}
-{\bfs y}_i^j)^2]\nonumber\\
&&\qquad +{T_0\over2}\sum_{j=1}^N\sum_{i=1}^{M}
[({\bfs x}_{i+1}^j-{\bfs x}_i^j)^2\phi_i^j+2{\bfs y}_i^{j2}-2{\bfs y}_i^j\cdot{\bfs y}_{i+1}^j\phi_{2i}^j]\ \equiv\ S_0(\phi)
\eea
To do this systematically, one can write 
$S\equiv S_0(\phi)+\Delta S(\phi)$ and treat $\Delta S(\phi)$
as a perturbation. Dropping $\Delta S$ is tantamount to the
mean field approximation. From this point of view $\phi$
{\it could} be chosen to be anything, but it {\it should} be chosen
to make the perturbative corrections as small as possible.
Dropping $\Delta S$ decouples the coordinate path integral from
the spin sum, so that the whole path integral factors into 
three parts ${\cal Z}={\cal Z}_x{\cal Z}_y{\cal Z}_s$,
or ${\cal F}={\cal F}_x+{\cal F}_y +{\cal F}_s$ with
\bea
e^{-{\cal F}_s}&=&\sum_{P_i^j=0,1}\exp\left\{-\sum_{ij}\left(\beta(1-P_i^j)
+(P_i^{j+1}-P_i^j)^2\ln g+\kappa_j^iP_i^j\right) 
\right\}=\prod_i{z}_s(\kappa_j^i)\\
{z}_s(\kappa_j)
&=&\sum_{P^j=0,1}\exp\left\{-\sum_{j}\left(\beta(1-P^j)+
(P^{j+1}-P^j)^2\ln g+\kappa_jP^j\right) 
\right\}
\eea
Specializing to static sources $\kappa_j=\kappa$,
the last sum can be thought of as the $N$th power of the
$2\times2$ matrix
\bea
T&=&\pmatrix{e^{-\beta}&ge^{-\kappa}\cr
ge^{-\beta}&e^{-\kappa}\cr}
\eea
which has eigenvalues
\bea
t_\pm=e^{-(\beta+\kappa)/2}\left[\cosh{\kappa-\beta\over2}
\pm\sqrt{\sinh^2{\kappa-\beta\over2}+g^2}\right]
\eea
The contribution of each eigenstate of $T$ to $z_s$
is weighted by $t_\pm^N$, and since $N\to\infty$ in the
continuum limit, the $+$ eigenstate will dominate:
\bea
\prod_iz_s(\kappa^i)\sim Ce^{N\sum_i\ln t_+(\kappa^i)}
\eea
Since the coordinate integrations are decoupled from the
spin sums at zeroth order in the mean field approximation, 
it follows (for static sources) that
\bea
\VEV{P_i^j}_0&=& -{\partial\over\partial\kappa^i}\ln t_+(\kappa^i)
\ =\ {1\over2}\left(1-{\sinh(\kappa_i-\beta)/2\over
\sqrt{\sinh^2(\kappa_i-\beta)/2+g^2}}\right)\\
\VEV{P_i^jP_{i+1}^j}_0&=& (\VEV{P_i^j}_0)^2
\eea
Then, making the choices $\phi_i^j=\VEV{P_i^j}_0\equiv \phi_i$
and $\phi_{2i}^j=\VEV{P_i^jP_{i+1}^j}_0=\phi_i^2$ optimizes
the perturbation theory in the sense that $\VEV{\Delta S}_0=0$.
These choices imply a linkage between $\phi_i$ and $\kappa_i$
which can be expressed as
\bea
\kappa_i&\approx&\kappa_{0i}(\phi)
\equiv\beta+2\ln\left\{{g(1-2\phi_i)
\over2\sqrt{\phi_i(1-\phi_i)}}
+\sqrt{1+{g^2(1-2\phi_i)^2\over4\phi_i(1-\phi_i)}}\right\}
\eea
We have added the 
subscript $0$ to $\kappa$ on the right of the last equation to
emphasize that that relation between $\kappa$ and $\phi$,
which we shall use to eliminate $\kappa$ in favor of
$\phi$ in the spin part of the effective action,
holds only at zeroth order and
neglects the back-reaction of the coordinate fields--the 
essence of the mean field approximation. 
The remaining two terms in ${\cal F}$ are determined
at zeroth order by
\bea
e^{-{\cal F}_x}&=&\int \prod_{ij}d{\bfs x}_i^j
\exp\left\{-{T_0\over2}\sum_{ij}\left[({\bfs x}_i^{j+1}
-{\bfs x}_i^j)^2+
\phi_i({\bfs x}_{i+1}^j-{\bfs x}_i^j)^2\right]\right\}\\
e^{-{\cal F}_y}&=&\int \prod_{ij}d{\bfs y}_i^j
\exp\left\{-{T_0\over2}\sum_{ij}\left[({\bfs y}_i^{j+1}
-{\bfs y}_i^j)^2+2(1-\phi_i^2){\bfs y}_i^{j2}+
\phi_i^2({\bfs y}_{i+1}^j-{\bfs y}_i^j)^2\right]\right\}
\eea
Here we restrict attention to the planar (with respect
to open string loops) 
evolution of the closed string,
so that periodic boundary conditions are appropriate.
Then by worldsheet translational invariance we can
expect that the mean field in the system
ground state is uniform over the
worldsheet $\phi_i^j=\phi$. It is then sufficient to
take a uniform source: $\kappa_j^i=\kappa$. We are also
interested in the continuum limit $M,N\to\infty$,
so the dominant contribution to the effective action,
${\cal A}={\cal F}-M(N+1)\kappa\phi$,
will be the bulk term $M(N+1){\cal V}$ proportional to the area.
The mean field will be determined by minimizing this term,
i.e. by minimizing the effective potential ${\cal V}(\phi)
={\cal V}_s+{\cal V}_x+{\cal V}_y$.
\begin{figure}
\begin{center}
\psfrag{phi}{$\phi$}
\psfrag{V_s}{$\hskip-12pt{\cal V}_s$}
\includegraphics[width=2.75in,height=3in]{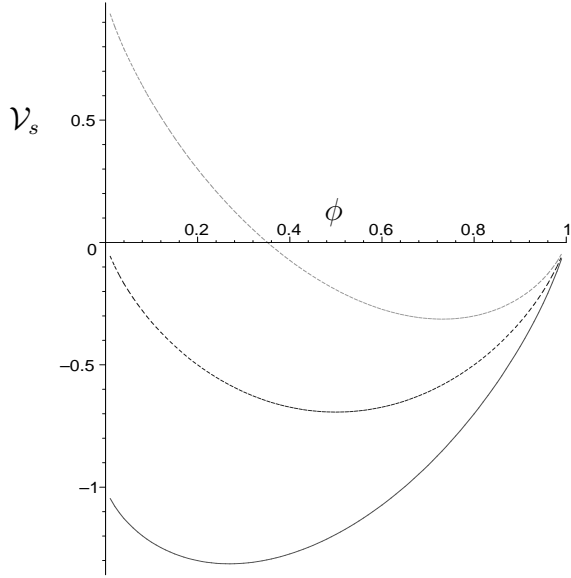}
\caption{The spin contribution to the effective potential, showing
the case of $g=1$ and $\beta=-1$ (lowest curve) $\beta=0$
(middle curve) and $\beta=1$ (highest curve).}
\label{vs1beta}
\end{center}
\end{figure}
We next list the three contributions to ${\cal V}$
in the mean field approximation, including
the $-\kappa\phi\approx-\kappa_0(\phi)\phi$ term 
in the spin contribution\footnote{
The technical details of their derivation can be gleaned
from the appendices of \cite{bardakcitmean,trantfishnet}.
Note that the condition $\kappa_0(\phi)=0$ determines 
a stationary point of ${\cal V}_s$ not ${\cal V}$. 
By minimizing
the total ${\cal V}$ we take  the back-reaction 
due to coordinate fluctuations into account
in an average way.}.
\bea
{\cal V}_s&=&{{\cal F}_s(\kappa_0(\phi))\over M(N+1)}
-\kappa_0(\phi)\phi
\ \to\ -\ln t_+-\kappa_0(\phi)\phi
\nonumber\\
&=&\beta(1-\phi)+(1-2\phi)\ln\left\{{g(1-2\phi)
\over2\sqrt{\phi(1-\phi)}}
+\sqrt{1+{g^2(1-2\phi)^2\over4\phi(1-\phi)}}\right\}\nonumber\\
&&-\ln\left\{{g
\over2\sqrt{\phi(1-\phi)}}
+\sqrt{1+{g^2(1-2\phi)^2\over4\phi(1-\phi)}}\right\}\\
{\cal V}_x&=&{{\cal F}_x\over M(N+1)} \to 
(D-2)\int_0^1dx \sinh^{-1}(\sqrt{\phi}\sin\pi x) = 
{2(D-2)\over\pi}\sum_{n=0}^\infty{(-)^n\phi^{n+1/2}\over
(2n+1)^2}\\
{\cal V}_y&=&{{\cal F}_y\over M(N+1)}\ \to\
(26-D)\int_0^1dx \sinh^{-1}\sqrt{{1\over2}(1-\phi^2)
+\phi^2\sin^2\pi x} 
\eea
Note that all of the dependence on the coupling $g$ and
the counterterm parameter $\beta$ is carried by
${\cal V}_s$. As an illustration of this dependence we
show a plot of ${\cal V}_s$ for $g=1$ and three different
values for $\beta$ in Fig.~\ref{vs1beta} and for
$\beta=0$ and three different values for $g$ in Fig.~\ref{vsg0}.
Plots of ${\cal V}_x$ and ${\cal V}_y$ 
are shown separately in Fig.~\ref{vxvy},
and combined (for the case $D=4$) in Fig.~\ref{vxy}. 
The non-monotonic
behavior of this last graph is a direct consequence of the
opposite monotonic behavior for Dirichlet and Neumann coordinates
evident in Fig.~\ref{vxvy}. 
These curves will be the same for all values of $g$ and
$\beta$. Of course the combined plot Fig.~\ref{vxy} will depend on
$D$ which controls the relative weight of the coordinates
with Neumann and Dirichlet boundary conditions.
Finally  in Figs.~\ref{a1beta}, \ref{ag0}, we plot the 
total effective potential ${\cal V}
={\cal V}_x+{\cal V}_y+{\cal V}_s$ for the same values
of $\beta$ and $g$ used in Figs.~\ref{vs1beta}, \ref{vsg0}.

In the mean field approximation, the
value of $\phi$ controls the effective tension of the string.
To see how, we write the effective action in the continuum
limit:
\bea
{\cal A}_{eff}&\to&{1\over2}\int d\tau d\sigma
\left[{\dot{\bfs x}}^2+{\dot{\bfs y}}^2+T_0^2(\phi{\bfs x}^{\prime2}
+\phi^2{\bfs y}^{\prime2})+{2\over a^2}(1-\phi^2){\bfs y}^2\right]
\eea
For oscillations parallel to the D-branes (in the ${\bfs x}$
directions) we have $T^{eff}_\parallel=T_0\sqrt{\phi}$,
and for oscillations perpendicular to the D-branes
we have $T^{eff}_\perp=T_0\phi$. The oscillations
perpendicular to the brane also have an effective mass
$m^{eff}=\sqrt{2(1-\phi^2)}/a$. For generic $\phi<1$
this mass is divergent in the continuum limit and would
have the effect of suppressing oscillations perpendicular to the
D-branes. For these oscillations to cost finite energy,
would require $\phi=1+O(a^2)$ as $a\to0$.

The plots of the effective potential in Figs.~\ref{a1beta},
\ref{ag0} all show that both endpoints $\phi=0,1$ are
local minima. The reason can be seen analytically from the endpoint
behavior of ${\cal V}$. In the $\phi\to0$ limit, ${\cal V}_x$
approaches zero with infinite positive slope, 
${\cal V}_y$ approaches its
limiting value with zero slope, and ${\cal V}_s$ approaches
0 with a negative infinite slope. 
\bea
{\cal V}^\prime_x\sim{D-2\over\pi\sqrt{\phi}}, \qquad
{\cal V}^\prime_y\sim0,\qquad{\cal V}^\prime_s\sim -\ln{1\over\phi},
\qquad {\rm for}~\phi\to0
\eea
Clearly, ${\cal V}^\prime_x$ dominates
as long as $D>2$, so ${\cal V}(\phi)$ rapidly increases
as $\phi$ increases from $0$. At the other endpoint, $\phi=1$,
${\cal V}_x$ approaches its value at finite slope, ${\cal V}_y$
approaches its value at negative infinite slope,
and ${\cal V}_s$ approaches its value at positive infinite
slope. 
\bea
{\cal V}^\prime_x\sim{D-2\over4}, \qquad
{\cal V}^\prime_y\sim-{26-D\over2\pi}\ln{1\over1-\phi},\qquad
{\cal V}^\prime_s\sim \ln{1\over1-\phi},
\qquad {\rm for}~\phi\to1
\eea
Here ${\cal V}_y,{\cal V}_s$ are comparable and we conclude
that
\bea
{\cal V}^\prime\sim -{26-D-2\pi\over2\pi}\ln{1\over1-\phi},
\qquad {\rm for}~\phi\to1,
\eea
and ${\cal V}$ will increase from its value as $\phi$ decreases
from 1 as long as $D<26-2\pi$.
It follows that, for small to moderate
coupling, only the endpoints $\phi=0,1$ are candidate
minima of the
effective potential. Which one is actually lower in energy 
is controlled
by the value of $\beta$, which is not known {\it a priori}
for general $g$.
From the free open string calculation we know
that $\beta\approx-3.91 +O(g^2)$ for $D=4$. For a value of $\beta$
this negative, $\phi=0$ is favored 
over $\phi=1$ by a very wide margin
for small to moderate couplings. Keeping $\beta$ fixed
at this value, one would have to go to $g>7$ for
a minimum with $0<\phi<1$ to develop.
Since the ``bare'' one
loop correction comes entirely from second order perturbation
theory, it should lower the zero coupling energy and hence
require the $O(g^2)$ contribution to $\beta$ to be positive,
making it a little less negative. 
In any case,
we can safely say that mean field theory predicts
that the effective string tension will vanish in the
system ground state at very weak coupling $g\ll1$.

For $g$ sufficiently large at fixed $\beta$, 
a local minimum in the effective
potential develops at some $\phi_0$ between 0 and 1.
However, at the critical coupling where
${\cal V}^{\prime\prime}(\phi_0)$ vanishes, this local minimum
has higher energy than one or both of the two endpoint
minima, so
this minimum initially describes a metastable phase. 
Near the critical point this metastable phase would
support finite energy spin waves, signifying the
emergence of a new Liouville-like degree of freedom
on the worldsheet. 
But eventually for larger coupling, this new minimum
could become a global minimum and
the system ground state would support a finite string tension. 
However, since we do not know $\beta$ for $g=O(1)$,
we cannot rule out the possibility that
$\beta$ becomes more and more negative as $g$ grows.
If this happens the local minimum just described
may never become a global one. Thus even in the
mean field approximation, the 
jury is still out on the question of whether
our lattice model of the bosonic string 
will actually support a finite string tension.

\section{Discussion and Conclusions}
In this article we have shown how to extend the
lattice worldsheet formalism 
for the bosonic string to allow for D-branes, and
we have applied a mean field approximation
to the resulting lattice model. Although the
mean field analysis allowed us to map out the possibile
phases of the system, our {\it a priori} ignorance of the
value of $\beta(g)$ leaves us uncertain about which
phase is actually realized when $g\geq O(1)$.
However, the mean field analysis is uneqivocal about the
weak coupling phase: in it the mean field $\phi=0$
and the string tension is quenched to zero.

There are many levels at which our results must
be regarded as provisional. First is the question: does the
lattice formalism accurately represent the
bosonic string perturbation theory? The answer is yes
only if the bulk and boundary counterterms we
have allowed for are sufficient to absorb all Lorentz
covariance violating artifacts due to the lattice cutoff.
The evidence for this is so far very meager. These
counterterms suffice to render the spectrum and tree scattering
amplitudes Lorentz covariant. At one loop, we have analyzed the
open string propagator and shown that for this specific one
loop process the counterterms suffice. It should not be too
difficult to check this conclusion for higher point
one loop amplitudes, but this has not yet been done.
We have no information on this issue at two loops and
beyond. Clearly more investigation of this quesiton is called for.

Even if the worldsheet lattice system is not Lorentz covariant
because of the need for more complicated counter-terms,
it remains a well-defined two dimensional system of
scalar fields interacting with an Ising spin system.
The physics of this system can be analyzed in its own right.
In this article we have begun this analysis within the
mean field approximation. The analysis suggests
that the system exhibits three distinct phases: a phase
with zero effective string tension ($\phi=0$), a phase
with maximal effective tension $T_0$ ($\phi=1$) and a
disordered spin phase, only stable at sufficiently
large coupling, with reduced effective tension
$T_0\sqrt{\phi}$ ($0<\phi<1$). This intermediate phase
could potentially support a meaningful infinite
tension limit in which the effective tension stays
finite. In such a limit the planar string diagrams
would go over into planar quantum field theory diagrams,
and one might gain insight into the large $N$ limit
of certain matrix field theories.
All of these conclusions
depend on the validity of the mean field approximation,
which by its very nature is somewhat dubious. 
But there are other approaches
to analyzing this system. In particular, Monte Carlo
methods seem particularly apt, since the path integrand is
positive definite and local. Such an analysis to test
the mean field conclusions would be very welcome.

Finally, we have to recognize that the tachyon in
the bosonic open string theory obscures the meaning
of the open string loop expansion our model is meant to
represent. By itself the tachyon could simply mean that
the system is being studied in an unstable vacuum,
and it might disappear once a stable vacuum is found.
Since our lattice model is a perfectly well-defined
physical system, its physics could provide information
about the correct stable ground state. Indeed, the
indication, from the mean field approximation
at weak coupling, that the string tension is
quenched to zero could be the ultimate fate of the
tachyon instability: the bosonic open string would then
be unstable to decaying into an infinite number
of string bits \cite{thornsakharov}.
If this is the case, the open bosonic string theory
would not be a good starting point for understanding
large $N$ gauge theory. However, the even G-parity
sector of the Neveu-Schwarz model is free of open string
tachyons and provides a more promising approach
to large $N$ QCD \cite{thornsubqcd,thornnonabelian}.
This possibility makes the extension of the 
lightcone lattice worldsheet formalism to include Grassmann
variables a particularly desirable next step. 

\vskip14pt
\noindent\underline{Acknowledgments}: 
I would like to thank Peter Orland and
Nati Seiberg for helpful discussions.
I would also like to acknowledge the
hospitality of the School of Natural Sciences
at the Institute for Advanced Study, where this
work was carried out.
This research was supported in part by the Ambrose Monell
Foundation and in part by the Department
of Energy under Grant No. DE-FG02-97ER-41029.

\end{document}